\documentclass[aip,rsi,preprint]{revtex4-1}

\usepackage{graphicx}
\usepackage{siunitx}
\usepackage{mhchem}
\usepackage{bm}
\usepackage{xcolor}

\usepackage[utf8]{inputenc}
\usepackage[T1]{fontenc}

\newcommand*{\vek}[1]{\bm{{#1}}}

\begin{document}

\title{Cryogenic frequency-domain electron spin resonance spectrometer based on coplanar waveguides and field modulation} 

\author{Bj{\"o}rn Miksch}\email[]{bjoern.miksch@pi1.physik.uni-stuttgart.de}\author{Martin Dressel}\author{Marc Scheffler}
\affiliation{1.\,Physikalisches Institut, Universität Stuttgart, Pfaffenwaldring 57, 70569 Stuttgart, Germany}

\date{\today}

\begin{abstract}
We present an instrument to perform frequency-domain electron spin resonance (ESR) experiments that is based on coplanar waveguides and field modulation. A large parameter space in frequency (up to \SI{25}{GHz}), magnetic field (up to \SI{8}{T}), and temperature (down to \SI{1.6}{K}) is accessible. We performed experiments on DPPH (2,2-diphenyl- 1-picrylhydrazyl) as a standard to calibrate the field modulation as well as on a carbon fibre sample to estimate the overall sensitivity of the instruments. Spectra of a ruby sample in a broad frequency- and field range at cryogenic temperatures are recorded with and without field modulation. The comparison reveals the improved signal-to-noise ratio achieved by field modulation.
\end{abstract}

\maketitle 

\section{Introduction}
\label{sec:Introduction}
Electron spin resonance (ESR) is the spectroscopic method of choice for studying magnetic properties of materials with unpaired electrons. An external static magnetic field leads to a Zeeman splitting between different energy levels of the spin system formed by unpaired electrons. Transitions between those levels can be detected as resonant absorption of an applied microwave irradiation. Information such as spin susceptibility, g-factor, and relaxation times can be extracted from the parameters of the absorption line like its intensity, resonance field, line-width and -shape\cite{poole_1983}.
Conventional ESR spectrometers operate at fixed microwave frequency utilising a resonant cavity for high sensitivity. Spectra are recorded by measuring the reflection from the cavity loaded with the sample whilst sweeping the external magnetic field.
This mode of operation has serious drawbacks when it comes to the investigation of materials with field-induced phase transitions or large zero-field splittings where even using multiple instruments at different fixed frequencies is not enough to resolve the whole phase diagram\cite{schwarze_2015,onose_2012,povarov_2011,schaufuss_2009,schwartz_2000}. 
To overcome this issue a number of magnetic resonance setups operating at multiple frequencies or even broadband have been developed so far. Planar one-dimensional resonators allow measurements at harmonics of their fundamental resonance frequency\cite{scheffler_2013,malissa_2013,ghirri_2015}. Broadband operation has been achieved with different approaches: tunable cavities\cite{schlegel_2010}, non-resonant coils\cite{mahdjour_1986}, coaxial lines\cite{rubinson_1989} or planar waveguides\cite{denysenkov_2003,giesen_2005,harward_2011,montoya_2014}. In previous studies we used coplanar transmission lines and demonstrated the possibility of frequency-domain ESR measurements in a large frequency range between 0.1 and \SI{67}{GHz}\cite{clauss_2013,wiemann_2015}.

Within this publication we describe an instrument capable of frequency-domain ESR measurements with magnetic field modulation allowing for a vastly improved sensitivity. While previous modulation-based on-chip studies focussed on thin-film ferromagnetic resonance at room temperature\cite{belmeguenai_2009,kalarickal_2006,chen_2018,beguhn_2012,maksymov_2015}, we are interested in a wider  parameter range, in particular low temperatures. We demonstrate the instrument's operation between 1.6 and \SI{300}{K} in a superconducting magnet allowing for fields up to \SI{8}{T}. Measurements of three different samples are performed to present the capabilities of the instrument. The stable free-radical molecule DPPH (2,2-diphenyl-1-picrylhydrazyl) with its narrow line-width is used as a standard to calibrate the amplitude of the magnetic field modulation. A single crystalline sample of ruby (\ce{Cr^3+}:\ce{Al2O3}) with a concentration of \SI{0.5}{\percent} is measured in a broad frequency- and temperature range as an example for a material with a more complex energy level scheme. Spectra are taken with as well as without field modulation to demonstrate the signal-to-noise level enhancement. The overall sensitivity of the instrument is estimated with an easy to handle carbon fibre sample\cite{herb_2018}.

\section{Experimental Setup}
\label{sec:Setup}
The setup of the frequency-domain ESR experiment consists of 3 parts: 
\begin{itemize}
  \item $^4$He cryostat with superconducting magnet and variable temperature insert
  \item ESR insert with probehead containing the sample box with coplanar waveguide and attached sample as well as the modulation coil
  \item Instrumentation for microwave generation and detection as well as driving the modulation field
\end{itemize}
The working principle of the individual parts is described in the following paragraphs  listing the devices used.

The instrument is set up in an Oxford Instruments Integra superconducting magnet in a $^4$He cryostat containing a variable temperature insert (VTI) within the magnet bore. Magnetic fields up to \SI{8}{T} can be applied in vertical direction parallel to the surface of the coplanar waveguide.
Temperatures as low as \SI{1.4}{K} can be achieved by running liquid helium through a needle valve into the sample space which is pumped by a mechanical pump. A Lakeshore Cernox CX-1050 Sensor is mounted in a bore in the copper sample stage directly below the sample box to ensure the accurate measurement of the sample temperature.  Two resistive heaters can be used to control the temperature via an Oxford Instruments ITC503 temperature controller; one is directly mounted on the sample stage in the ESR insert and used primarily for higher temperatures above \SI{10}{K}, the other one is mounted on the heat exchanger at the needle valve of the VTI allowing for precise control in the low-temperature region.

\begin{figure}
\includegraphics[width=8cm]{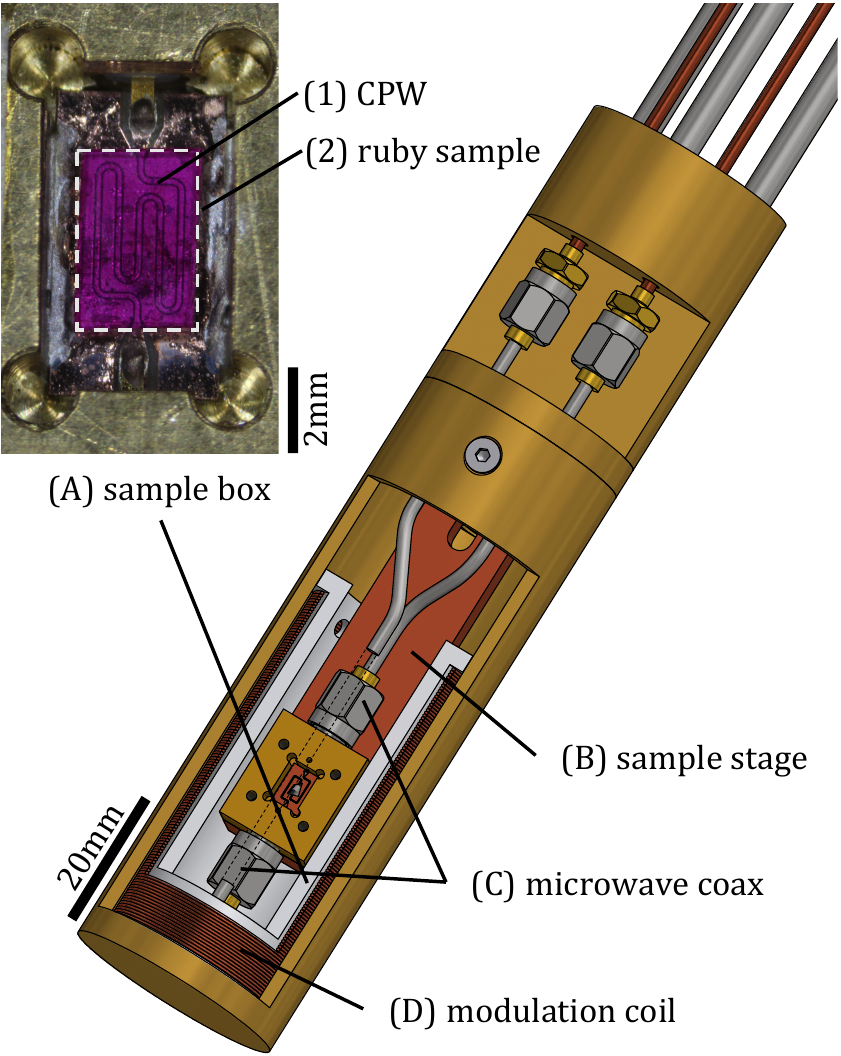}
\caption{\label{fig:probehead}
Cutaway view of the probehead at the lower end of the ESR insert. (A) Enlarged photo of a sample box with a mounted and contacted coplanar waveguide (1) and an attached ruby sample (2). (B) The copper sample stage contains a resistive heater and a temperature sensor. (C) Microwave coaxial cables for transmission measurement. (D) Modulation coil wound on teflon tube.
}
\end{figure}

From top to bottom the ESR insert is built up as follows: A KF flange seals the VTI exchange gas chamber. It has vacuum feedthroughs for the two 0.085 inch semi-rigid coaxial cables and a hermetically sealed connector for the wires to the thermal sensor and heater as well as the wires carrying the modulation current. 
The probehead located in the centre of the magnet bore is attached to the flange via three thin walled tubes which incorporate the wiring.
Figure \ref{fig:probehead} shows a cutaway view of the construction of the probehead. The sample stage which is manufactured from copper incorporates the thermal sensor and a \SI{40}{\ohm} resistive cartridge heater. Within the probehead two short $\sim$\SI{10}{cm} segments of semi-flexible coaxial cable are used as a connection between the sample box and the semi-rigid cables within the insert to allow for different sizes of sample boxes. 
The modulation coil is wound on a PTFE tube with a length of \SI{60}{mm} and a radius of \SI{13}{mm}. It is attached to the inside of the cylindrical cap covering the probehead. This allows for sliding the coil over the whole assembly after mounting the sample box and ensures the sample box to be centred within the coil. 

The coplanar waveguides (CPWs) are fabricated from \SI{1}{\micro\meter} copper on R-cut \SI{430}{\micro\meter} sapphire substrates\cite{javaheri_2016}. Unless stated differently, for the data shown in this publication the follwing dimensions are used: \SI{100}{\micro\meter} inner conductor width $W$ and \SI{42}{\micro\meter} gaps $S$ between inner conductor and ground planes for a nominal \SI{50}{\ohm} impedance. The cross-section of a coplanar waveguide is sketched in figure \ref{fig:carbon}. A microwave magnetic field $\vek{H_1}$ perpendicular to the static external field $\vek{H_0}$ is required in the regular ESR geometry. As the microwave magnetic field encircles the inner conductor of the CPW and is therefore always perpendicular to the direction of the waveguide, the standard ESR geometry is obtained by orienting the CPW parallel to the static magnetic field $\vek{H_0}$. To increase the effective interaction volume of the microwaves with a larger sample, the waveguide can be in meander shape as shown in figure \ref{fig:probehead}(A) where the five vertical segments of the CPW are responsible for the main part of the signal. The field distribution above the CPW depends on its geometrical parameters where the width of the inner conductor defines the length scale for the roll-off of the microwave magnetic field\cite{simons_2001}. Waveguides of smaller width can be beneficial for thin film samples as the microwave magnetic field is confined to a smaller volume above the surface and thus is of larger magnitude. Mounting bulk samples on the waveguide always leads to a small unavoidable gap between the two\cite{ebensperger_2019}. Waveguides of larger dimensions, allowing the field to penetrate deeper into the sample volume, can thus be beneficial.

The CPW is mounted into a sample box fabricated from brass. The transition from the coaxial cables to the CPW is realised using sparkplug launchers attached to the sample box. The samples are attached to the CPW via Apiezon N cryogenic vacuum grease.
Thermal contact to the sample stage is ensured by the grease as well as the helium exchange gas within the probehead.

\begin{figure}
\includegraphics[width=8cm]{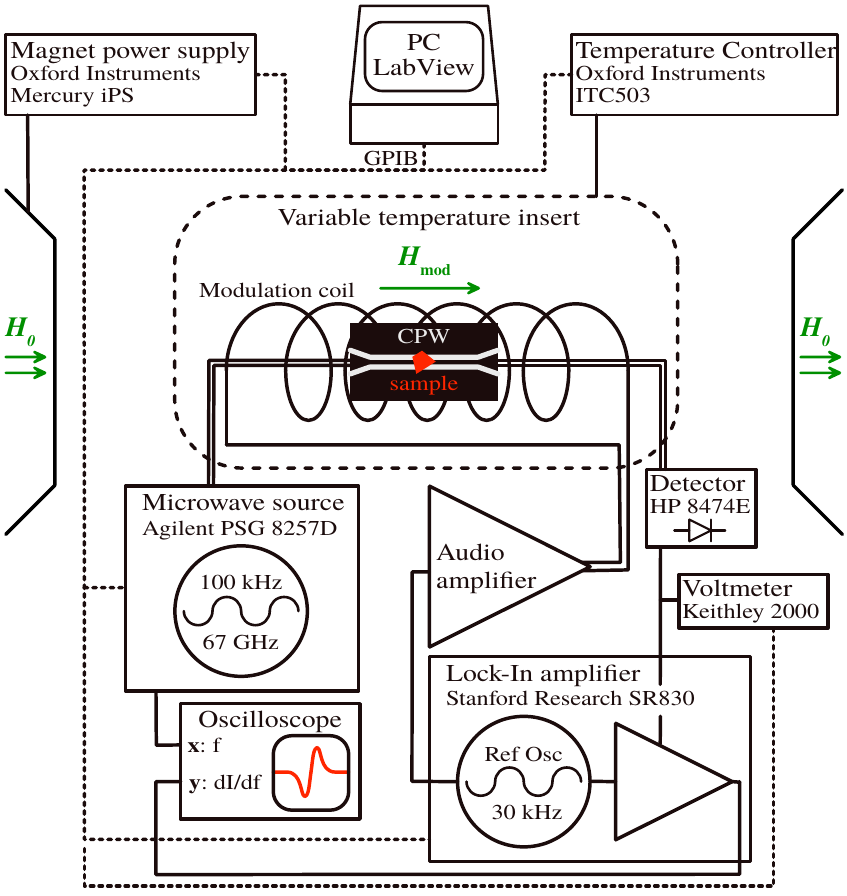}
\caption{\label{fig:setup}
Schematic description of the instrumentation. The magnet power supply and temperature controller are used to set the external parameters magnetic field and sample temperature. The frequency-swept microwave signal is generated by an analog signal generator and detected by means of a diode detector. The lock-in amplifier is used to generate the modulation signal as well as to demodulate and record the spectra. An audio amplifier drives the modulation coils. All instruments are computer controlled by a LabView program running an automated measurement sequence. It allows to set all the external parameters, control the microwave measurements, and save the spectra.
}
\end{figure}

Figure \ref{fig:setup} shows a schematic view of the instrumentation of the setup which is controlled by a PC running LabVIEW. The equipment used for the phase-sensitive detection scheme employed for the ESR measurements can be divided in three functional groups: microwave generation, detector and lock-in amplifier, and modulation amplifier. The microwave signal is generated by an Agilent PSG 8257D analog signal generator in the frequency range between \SI{100}{\kilo\hertz} and \SI{67}{\giga\hertz} with an output power between \SI{-20}{dBm} and \SI{15}{dBm}. External power levelling can be used to compensate for changes in the microwave power. In this case a detector diode for levelling is built in the setup with a directional coupler at the output of the insert. To avoid distortion of the ESR signal, the bandwidth of the automatic levelling control circuit of the microwave generator is limited to a frequency of \SI{1}{kHz} and thus much lower than the modulation frequency of \SI{30}{kHz}. 

The transmission through the insert is measured with a diode detector (HP 8474E). The voltage from the video output of the detector is directed into a digital lock-in amplifier (Stanford Research SR830) which also generates the sinusoidal modulation signal. A voltmeter (Keithley 2000) can be used to measure the effective value of the transmission signal from the detector directly. The modulation coil is driven by an audio power amplifier which is fed with the modulation signal from the lock-in amplifier. The frequency swept ESR signal is recorded digitally within the lock-in amplifier and can be directly read out via GPIB with the controlling PC.

Modulating the magnetic field not only leads to a modulated ESR absorption but also always induces an undesired voltage within the CPW. This induced voltage is amplified by the lock-in amplifier as it oscillates with the modulation frequency. It occurs always \SI{90}{\degree} out of phase with the modulation field whereas the ESR signal is modulated in phase. With the microwave frequency being detuned from the ESR resonance, only the induced voltage is detected by the lock-in amplifier and hence it can be used to adjust the phase of the detection channels to have the ESR signal solely in the Y channel of the lock-in amplifier with correct phase.

\section{Instrument Performance}
\label{sec:Performance}
\begin{figure}
\includegraphics[width=8cm]{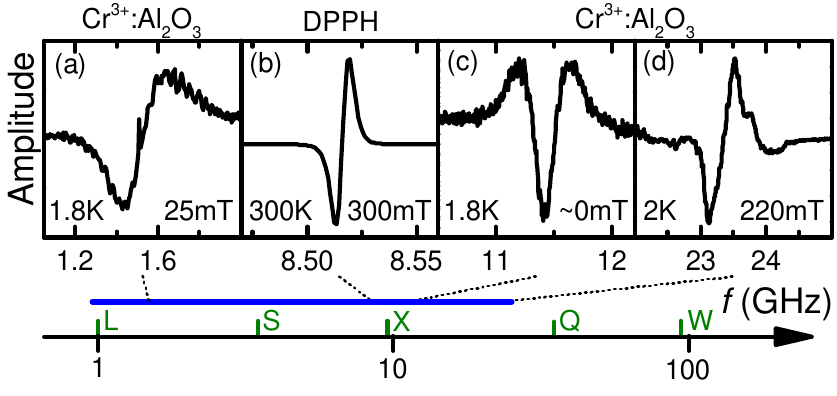}
\caption{\label{fig:freq_scale}
Demonstration of the presented instrument's parameter range. (a) and (d) spectra of different transitions in the ruby sample at low temperature, (b) ESR of DPPH at room temperature, (c) direct measurement of the zero-field splitting in ruby.
}
\end{figure}

A set of spectra is shown in figure \ref{fig:freq_scale} to demonstrate the capabilities of the instrument. Spectra (a), (c) and (d) are recorded at low temperature and show transitions in the ruby sample at different combinations of frequency and magnetic field; here (c)  demonstrates the benefit of the frequency-domain operation allowing for the direct observation of the zero-field splitting. Spectrum (b) shows the operation around $g=2$ with the measurement of a DPPH standard sample at room temperature.

\begin{figure}
\includegraphics[width=8cm]{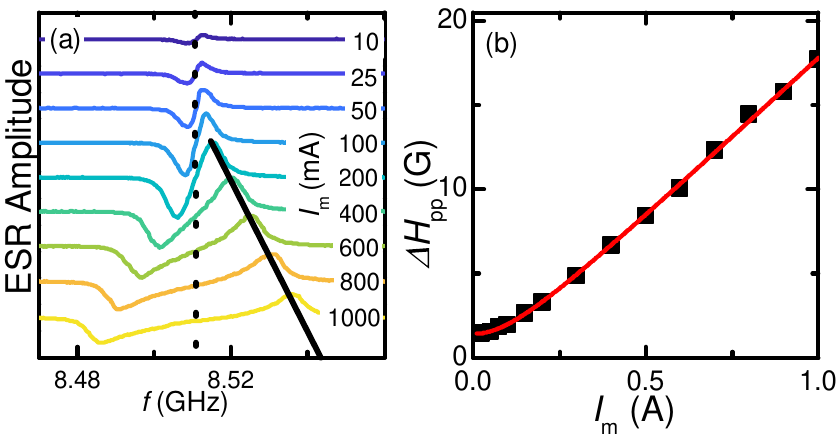}
\caption{\label{fig:modamp}
Calibration of the modulation amplitude by overmodulating the DPPH ESR signal. (a) Spectra recorded for different modulation amplitudes at $H_0=\,$\SI{300}{mT}. (b) Fitting the peak-to-peak line-width gives a value of \SI{0.019}{G\per mA} for the modulation amplitude $H_m$ as well as the true line-width $\Delta H_{pp}=\,$\SI{1.4}{G}. 
}
\end{figure}

Calibration of the modulation amplitude is done by recording overmodulated spectra of a DPPH powder sample. Spectra recorded with a modulation current between \SI{10}{mA} and \SI{1000}{mA} are shown in figure \ref{fig:modamp}(a). Extracting the peak-to-peak line-width in frequency-domain and converting it to magnetic field units with $g_\text{DPPH}=2.0036$ leads to the line-width $\Delta H_{pp}$ shown in figure \ref{fig:modamp}(b). Fitting the model for the overmodulated Lorentzian line from [\onlinecite{poole_1983}] to the data results in a value of \SI{0.019}{G\per mA} and thus a maximal modulation amplitude of \SI{19}{G} at a current of \SI{1}{A} through the coil -- a value at which the cryostat is still able to keep the temperature below \SI{1.6}{K}. The possible modulation amplitude is thus comparable to the one in a standard cavity for a commercial X-Band ESR spectrometer, e.g. \SI{32}{G} in a Bruker ER4102ST rectangular cavity.

\begin{figure}
\includegraphics[width=8cm]{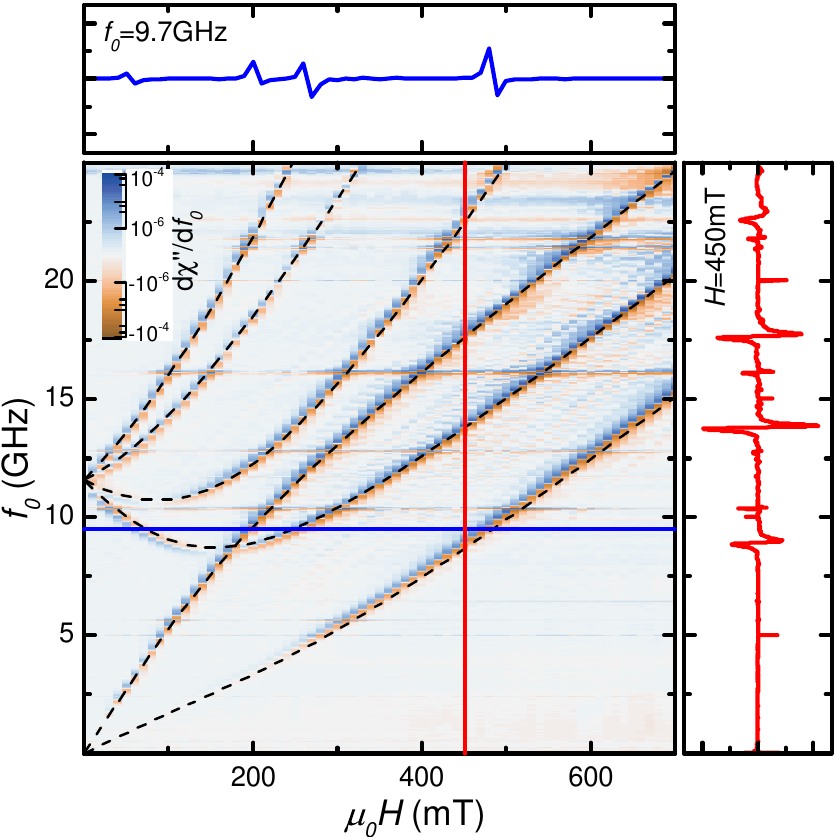}
\caption{\label{fig:ruby}
Frequency-domain ESR measurement of \ce{Cr^3+}:\ce{Al2O3} at $T=2\,$K. An exemplary spectrum at \SI{450}{mT} is shown in red on the right. A field-domain spectrum can be extracted as shown in blue for the X-Band frequency of \SI{9.7}{GHz}. Dashed lines show the 6 predicted transitions between the 4 levels of the $S=\frac{3}{2}$ system for an angle of $\theta=78^{\circ}$ in between the static field and the c-axis of the crystal.
}
\end{figure}

The broadband operation of the instrument is demonstrated by the measurement of the ruby sample shown in figure \ref{fig:ruby}. Individual frequency-domain spectra are recorded at increasing magnetic field in steps of \SI{10}{mT}. One exemplary spectrum for $H_0=\SI{450}{mT}$ is shown in the right graph in red. The frequency is swept in the range between 0.1 and \SI{25}{GHz}. Here 71 spectra are recorded to sample the whole range between 0 and \SI{700}{mT}. The extracted field-domain spectrum at the X-Band frequency of \SI{9.7}{GHz} is shown in blue in the top graph. Field modulation is set to an amplitude of \SI{1.9}{G} at a frequency of \SI{30}{kHz} and the spectra are recorded with \SI{100}{ms} integration time of the lock-in amplifier. All six transitions between the four levels of the $S=\frac{3}{2}$ spin of the \ce{Cr^3+} ions are visible in the whole spectrum down to a frequency of \SI{1}{GHz}\cite{geusic_1956,schulzdubois_1959,schuster_2010}. The dashed lines show the resonance spectrum calculated with the spin Hamiltonian from [\onlinecite{chang_1978}]. For an angle of $\theta=78^{\circ}$ between the static magnetic field and the cylindrical symmetry axis of the \ce{Cr^3+} Hamiltonian -- the c-axis of the \ce{Al2O3} crystal -- the calculated spectra coincide well with the measured ones. The horizontal stripes visible in the graph presumably originate from undesired resonant modes within the sample box or the microwave transmission line\cite{wiemann_2015}.

\begin{figure}
\includegraphics[width=8cm]{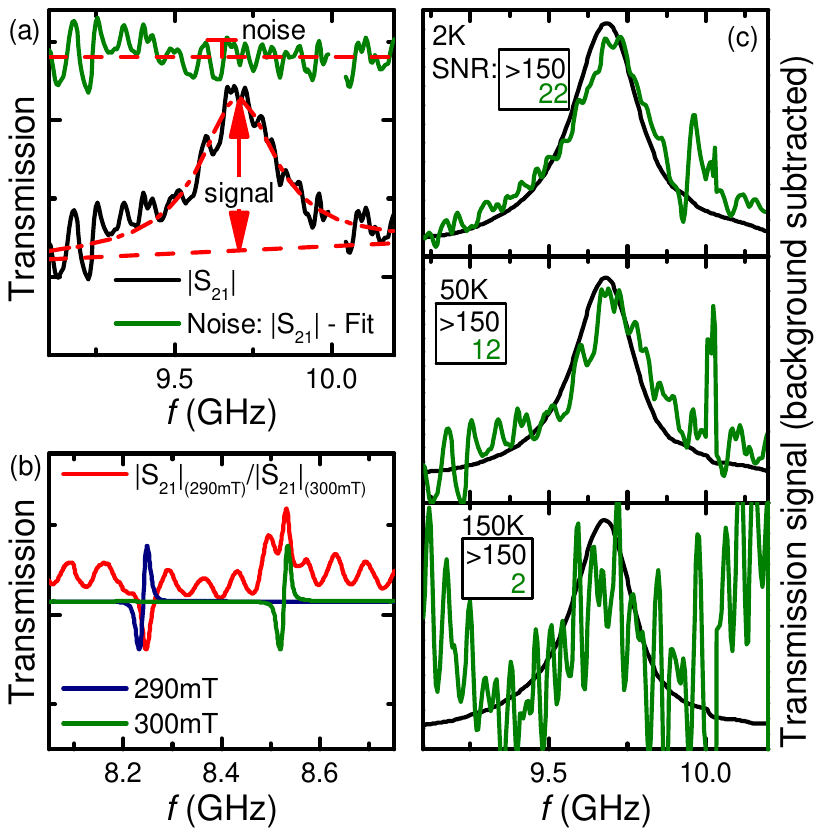}
\caption{\label{fig:SNR}
(a) Absorption spectrum and its noise floor. $\text{SNR}=A_\text{Signal}/2\sigma_\text{Noise}$. (b) ESR spectra of DPPH at \SI{300}{K}. Red: transmission spectrum divided by spectrum at \SI{10}{mT} higher field to reduce background. Blue and green: corresponding spectra recorded with field modulation. (c) $m_S=+\frac{1}{2}$ to $m_S=+\frac{3}{2}$ transition in ruby at $H=\SI{200}{mT}$; green: transmission signal, black: transmission obtained from field-modulated signal by integration.
}
\end{figure}

In figure \ref{fig:SNR} the improvements achieved by field modulation are shown for two examples. The microwave transmission signal $|S_{21}|$ always carries a standing wave pattern due to impedance mismatches at the microwave connections contributing to the background of the transmission signal. This background can be seen in the ESR spectrum of DPPH shown in figure \ref{fig:SNR}(b). Here the background has already been reduced by dividing the signal at \SI{290}{mT} by the transmission at \SI{10}{mT} higher field. As the pattern slightly shifts upon applying magnetic field the background cannot be completely removed by this procedure. Field-modulation allows to effectively suppress any static background from the measurements as seen by the two spectra shown for $H=\SI{290}{mT}$ and \SI{300}{mT}. 

To demonstrate the reduction of the noise level, field-modulated and pure transmission spectra have been recorded simultaneously. The noise floor is extracted from the data by fitting the Lorentzian absorption and subtracting the fit from the data. The ratio of the signal amplitude and the standard deviation of the noise floor is taken as a measure of the signal-to-noise ratio (SNR) as demonstrated in figure \ref{fig:SNR}(a). Spectra of a selected transition in the ruby sample for three different temperatures are compared in figure \ref{fig:SNR}(c). The signal-to-noise ratio for the transmission spectra decreases with increasing temperature. At \SI{150}{K} the signal can hardly be distinguished from the noise floor anymore. In comparison the transmission obtained by integrating the field-modulated spectra shows almost no noise within the instrument's resolution even at \SI{150}{K}.

\begin{figure}
\includegraphics[width=8cm]{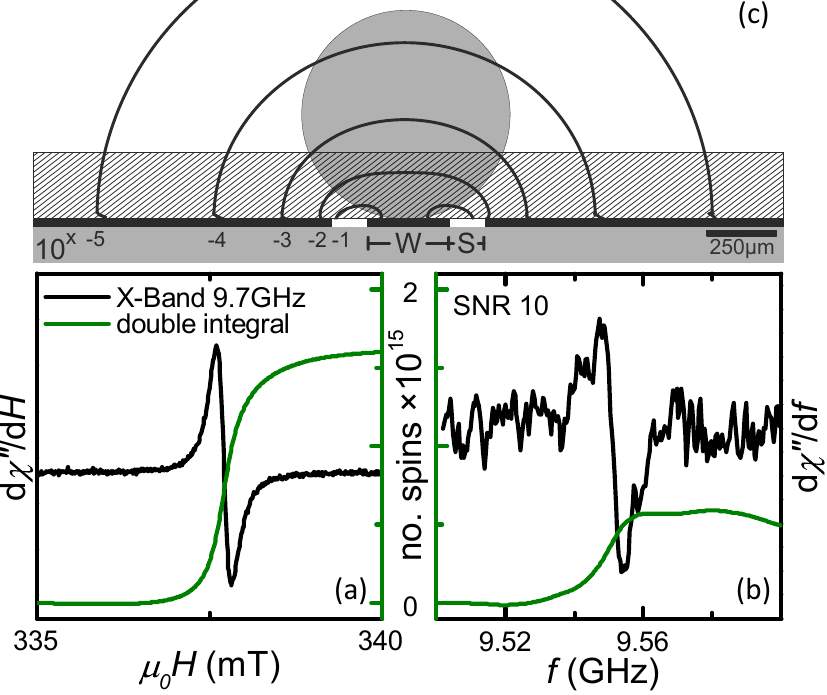}
\caption{\label{fig:carbon}
Room-temperature ESR spectra of a carbon fibre reinforced polymer rod (diameter $\times$ length $\SI{0.75}{mm}\times\SI{1.5}{mm}$). (a) Field-swept spectrum obtained with a conventional X-Band spectrometer using a standard rectangular $\text{TE}_{102}$ cavity. The number of spins in the sample is determined to be $1.6\times10^{15}$ by double integration. (b) Frequency-swept spectrum recorded with the presented instrument at \SI{337}{mT}. About one third of the whole number of spins contribute to the spectrum. (c) Dimensions of the sample compared to the waveguide with inner conductor width $W$ and gap size $S$. Contour lines indicate the decrease of the microwave magnetic field amplitude $H_1$ (in powers of 10) when moving away from the waveguide.
}
\end{figure}

To estimate the number of spins needed to detect a signal at room temperature, a carbon fibre sample with a line width of \SI{2}{G} has been used (Conrad SE carbon reinforced polymer rod with \SI{0.75}{mm} diameter). The number of spins in the sample was determined with a commercial X-Band spectrometer to be $1.6\times 10^{15}$ by double integration and calibration with a $\gamma$-irradiated Alanine standard as shown in figure \ref{fig:carbon}. The frequency-domain measurement was performed on a waveguide with \SI{300}{\micro\meter} inner conductor width $W$ to probe a larger amount of the sample. The probed volume of the sample can be calculated from the absolute value of the microwave magnetic field $\vek{H_1}$ above the waveguide according to [\onlinecite{simons_2001}]. The ESR signal is proportional to $|\vek{H_1}|^2$. Integrating over the lateral dimensions shows that \SI{90}{\percent} of the ESR signal originates from the volume up to a height of $0.8\cdot W$ above the waveguide as indicated by the hatched area in the sketch in figure \ref{fig:carbon} (c). For the geometries of the experiment with $W=\SI{300}{\micro\meter}$ and the sample diameter of \SI{750}{\micro\meter}, approximately \SI{30}{\percent} of the sample volume contributes to the signal. Therefore $5\times 10^{14}$ spins yield the signal shown in figure \ref{fig:carbon} (b). With \SI{1}{mW} of incident microwave power, \SI{30}{ms} integration time and a modulation amplitude of \SI{1}{G} a signal-to-noise ratio of about 10 is reached. By increasing the integration time a sensitivity of the instrument in the order of $10^{13}$ spins per \SI{1}{G} line-width is within reach for frequencies in the X-Band. Despite the the larger versatility of this instrument, the sensitivity is only 2 orders of magnitude lower than the $10^{11}$ spins per \SI{1}{G} line-width sensitivity specification of the Bruker EMXplus conventional fixed-frequency X-Band ESR spectrometer with Bruker ER4102ST standard rectangular $\text{TE}_{102}$ cavity resonator\cite{eaton_2010}.

\section{Conclusion}
\label{sec:conclusion}
We have presented an instrument to perform frequency-domain ESR measurements with field modulation. Operation of the instrument between 1.6 and \SI{300}{K} in the frequency range from 0.1 to \SI{25}{GHz} and magnetic fields up to \SI{700}{mT} has been demonstrated with measurements of a ruby sample. The frequency-domain operation allows to track all 6 transitions of the $S=\frac{3}{2}$ system and to directly observe the zero-field splitting. Introducing the field modulation leads to a vastly improved signal-to-noise ratio as shown by the temperature dependent experiments on ruby. The amplitude of the field modulation was calibrated with experiments on DPPH. An overall sensitivity of the instrument in the range of $10^{13}$ spins at room temperature is within reach as shown by measurements of a carbon fibre sample.

\begin{acknowledgments}
We thank G. Untereiner and M. Ubl for experimental support and G. G. Lesseux for fruitful discussions. Financial support by the Deutsche Forschungsgemeinschaft (DFG), grants SCHE 1580/2 and SCHE 1580/6, is thankfully acknowledged.
\end{acknowledgments}

\bibliography{broadband_esr.bib}

\end{document}